\documentclass[iopams,aps,ams,12pt]{iopart}
\expandafter\let\csname equation*\endcsname\relax
\expandafter\let\csname endequation*\endcsname\relax

\usepackage{epsf,latexsym,graphicx,amsmath}

\begin{document}

\def \ba {\begin{eqnarray}}
\def \ea {\end{eqnarray}}
\def \vk {\mathbf{k}}

\def \Bi2212 {Bi$_2$Sr$_2$CaCu$_2$O$_{8+\delta}$}
\def \LSCO {La$_{2-x}$Sr$_{x}$CuO$_{4}$}
\def \YBCO {YBa$_{2}$Cu$_3$O$_{6.6}$}

\title[Angle and frequency dependence of self-energy from spin fluctuations]
{Angle and frequency dependence of self-energy from spin
fluctuations mediated $d$-wave pairing for high temperature
superconductors }

\author{Seung Hwan Hong$^1$ and Han-Yong Choi$^2$ }

\address{$^1$ Department of Physics and Institute for Basic Science
Research, SungKyunKwan University, Suwon 440-746, Korea}

%\footnote{hychoi@skku.ac.kr: To whom the correspondences should be addressed.}

\address{$^2$ Department of Physics and Institute for Basic Science
Research, SungKyunKwan University, Suwon 440-746, Korea, Asia
Pacific Center for Theoretical Physics, Pohang 790-784, Korea}
\ead{hychoi@skku.ac.kr}

\begin{abstract}

We investigated the characteristics of the spin fluctuations
mediated superconductivity employing the Eliashberg formalism. The
effective interaction between electrons was modeled in terms of
the spin susceptibility measured by the inelastic neutron
scattering experiments on single crystal \LSCO \ superconductors.
The diagonal self-energy and off-diagonal self-energy were
calculated by solving the coupled Eliashberg equation
self-consistently for chosen spin susceptibility and
tight-binding dispersion of electrons. The full momentum and
frequency dependence of the self-energy is presented for the
optimal, overdoped, and underdoped LSCO cuprates in
superconductive state. These results may be compared with the
experimentally deduced self-energy from ARPES experiments.

\end{abstract}

\pacs{PACS: 74.20.-z, 74.25.-q, 74.72.Gh }

\submitto{\JPCM}

%\keywords{Eliashberg equation, d-wave pairing, spin fluctuations, cuprate superconductor}

\maketitle

%%%%%%%%%%%%%%%%    Introduction   %%%%%%%%%%%%%%%%%%%%%%%%%%%

\section{Introduction}

One of the leading contenders for the $d$-wave pairing mechanism
of cuprate superconductors is the spin fluctuations.
Superconductivity (SC) mediated by the spin fluctuations has a
long history.\cite{Scalapino12rmp,Monthoux07nature,Norman06arxiv}
In support of the spin fluctuation mechanism,
Scalapino,\cite{Scalapino12rmp} noticing the commonalities among
the heavy fermion, cuprate, and Fe superconductors, argued that
(a) Their chemical and structural makeup, their phase diagrams,
and the observation of a neutron scattering spin resonance in the
superconducting phase support the notion that they form a related
class of superconducting materials. (b) A number of their observed
properties are described by Hubbard-like models. (c) Numerical
studies of the effective pairing interaction in the Hubbard-like
models find unconventional pairing mediated by an $S = 1$
particle-hole channel. He proposed that spin fluctuation mediated
pairing provides the common thread which is responsible for
superconductivity in all of these materials.

Along the same line have many works been published. Noteworthy is
the work by Dahm $et~al.$\cite{Dahm09naturephys} They measured the
spin susceptibility from inelastic neutron scattering (INS)
experiments on \YBCO\ and used it as the effective interaction
(the Eliashberg function) between electrons to calculate the
diagonal self-energy from the Eliashberg equation. The calculated
spectral function produced similar results as the measured angle
resolved photoemission spectroscopy (ARPES) intensity from the
same \YBCO\ crystal. They claimed that a self-consistent
description of ARPES and INS can be obtained within the
Eliashberg formalism for the cuprates (by adjusting a single
parameter, the fermion-spin coupling strength.) Their work,
however, did not take full consideration of the frequency and
momentum dependence of the diagonal and off-diagonal self-energy.
This point is crucial in that, for instance, the momentum
dependence of the peak position of the self-energy which is one
of the ongoing discussions in the field can only be addressed by
calculations without assuming {\it ad hoc} momentum dependence.
See the remarks in Sec.\ 5 below.

Here, we revisit this spin fluctuation scenario by computing the
angle, i.e., the momentum direction in the Brillouin zone (BZ),
and frequency dependence of the diagonal, $\Sigma(\vk,\omega)$,
and off-diagonal self-energy, $\phi(\vk,\omega)$. The cause of the
angle and frequency dependence can provide an important clue
about the pairing mechanism.\cite{Anderson07science,Maier08prl}
%Recall that pairing is a Fermi surface (FS) instability in the
%normal state, that is, the pairing interaction is already present
%in the normal state. Therefore, the putative pairing interaction
%should be able to account for other physical properties than the
%pairing in the normal as well as SC states.\cite{Choi12jkps}
The diagonal self-energy is also called normal self-energy
(``normal'' here means the particle-hole channel and should not
be confused with the ``normal'' as in the normal state meaning
above $T_c$), and off-diagonal self-energy is also called
anomalous or pairing self-energy. We employ the phenomenological
fermion-spin coupling.\cite{Monthoux07nature,Dahm09naturephys}
 \ba
H_{fs} = \alpha \sum_{k,q,a,b} c^\dag_{k+q,a} c_{k,b}
\sigma_{a,b},
 \ea
where $\alpha$ is the coupling strength of the dimension of
energy, and $c$ and $\sigma$ are the fermion and spin operators.
Although the Eliashberg formalism is not firmly established for
spin fluctuation mediated superconductivity, perhaps, our resort
is that the ratio $\lambda \omega_c /E_F$ is $ < 1$. $\lambda$ is
the dimensionless coupling constant, $\omega_c$ the cutoff of the
spin fluctuation frequency, and $E_F$ is the Fermi energy. Also,
Millis argued in Ref.\ \cite{Millis92prb}, in justifying the
numerical Eliashberg approaches, that the $d$-wave
superconductivity induced by antiferromagnetic (AF) spin
fluctuations is due essentially to high-energy part and not to
the strong low-lying AF fluctuations producing the mass
enhancement and scattering.

For the Eliashberg function, $\alpha^2 F({\bf q},\omega)$, we
take like Dahm $et ~al.$ the imaginary part of the spin
susceptibility, $\chi_{sp} ({\bf q},\omega)$, measured from INS.
High quality INS data require large size single crystals and the
INS data over wide momentum and energy range are mainly from YBCO
or LSCO compounds. A functional form of the spin susceptibility
obtained by fitting the INS data is given in the literature for
optimally doped (OP) \LSCO $~(x=0.16,~ T_c =38.5$ K) by Vignolle
$et~al.$\cite{Vignolle07naturephys} and for overdoped (OV),
($x=0.22,~ T_c =26$ K),\cite{Lipscombe07prl} and underdoped (UD)
\LSCO \ ($x=0.08,~ T_c =22$ K)\cite{Lipscombe09prl} by Lipscombe
$et~al.$\cite{Lipscombe08thesis} The diagonal and off-diagonal
self-energy and the quasi-particle (qp) energy shift,
$X(\vk,\omega)$, are computed self-consistently for OP, OV, and
UD \LSCO \ using the INS measured spin susceptibility. See Eqs.\
(\ref{self-en2}) and (\ref{spectral-fun}) below.

In the following section 2, we present the Eliashberg formalism
used to calculate the self-energy from the given spin
susceptibility spectrum, $\chi_{sp} ({\bf q},\omega)$. Some
preliminary analysis for the energy scales of the self-energy is
given in section 3 before presenting results of numerical
calculations. In section 4, the numerical results are presented
for OP, OV, and UD \LSCO \ focusing on the angle dependence of
the position and intensity of the peaks in the self-energy. There
are two sources for the peaks in the self-energy in SC state: the
peaks in the density of states (DOS) and the spin susceptibility.
We will discuss how the two between them show up in the
self-energy. The summary and outlooks will follow in section 5.

%%%%%%%%%%%%%%%%       Theory      %%%%%%%%%%%%%%%%%%%%%%%%%%%

\section{Formalism}

The $d$-wave Eliashberg equation is given
by\cite{Sandvik04prb,Yun11prb}
 \ba \label{eliashberg}
\widetilde{\Sigma}(\mathbf{k},\omega) &=&
\int^{\infty}_{-\infty}d\epsilon\int^{\infty}_{-\infty}d\epsilon'
S \sum_{\mathbf{k'}}
A_{S}(\mathbf{k'},\epsilon)\alpha^{2}F^{(+)}(\mathbf{k},\mathbf{k'},\epsilon'),
 \nonumber \\
X(\mathbf{k},\omega)&=&\int^{\infty}_{-\infty}d\epsilon\int^{\infty}_{-\infty}d\epsilon'
S \sum_{\mathbf{k'}}
A_{X}(\mathbf{k'},\epsilon)\alpha^{2}F^{(+)}(\mathbf{k},\mathbf{k'},\epsilon'),
 \nonumber \\
\phi(\mathbf{k},\omega)&=&
-\int^{\infty}_{-\infty}d\epsilon\int^{\infty}_{-\infty}d\epsilon'
S \sum_{\mathbf{k'}}
A_{\phi}(\mathbf{k'},\epsilon)\alpha^{2}F^{(-)}(\mathbf{k},\mathbf{k'},\epsilon'),
 \nonumber \\
S &=&
\frac{f(\epsilon)+n(-\epsilon')}{\epsilon+\epsilon'-\omega-i\delta},
 \ea
where $f(\omega)$ and $n(\omega)$ are the Fermi and Bose
distribution functions, respectively. $\widetilde{\Sigma}$ is the
symmetric part of the diagonal self-energy $\Sigma$, $X$ the shift
of the qp dispersion, and $\phi$ is the off-diagonal self-energy.
The diagonal and off-diagonal Eliashberg functions,
$\alpha^{2}F^{(+)}$ and $\alpha^{2}F^{(-)}$, are given by
 \ba
 \label{eliash-fun}
\alpha^{2}F^{(+)}(\mathbf{k},\mathbf{k'},\epsilon') =
\alpha_{ch}^{2}(\mathbf{k},\mathbf{k'})
F_{ch}(\mathbf{k}-\mathbf{k'},\epsilon') +
\alpha_{sp}^{2}(\mathbf{k},\mathbf{k'})
F_{sp}(\mathbf{k}-\mathbf{k'},\epsilon'), \nonumber \\
 \alpha^{2}F^{(-)}(\mathbf{k},\mathbf{k'},\epsilon') =
\alpha_{ch}^{2}(\mathbf{k},\mathbf{k'})
F_{ch}(\mathbf{k}-\mathbf{k'},\epsilon') -
\alpha_{sp}^{2}(\mathbf{k},\mathbf{k'})
F_{sp}(\mathbf{k}-\mathbf{k'},\epsilon'),
 \ea
where the subscripts $ch$ and $sp$ refer to the channels by which
the bosonic modes transform. For instance, the charge
fluctuations belong to the $ch$ channel, and the spin and current
fluctuations to the $sp$ channel. The various spectral functions
are given by
 \ba
 \label{spectral-fun}
A_{S}(\mathbf{k},\omega)=
-\frac{1}{\pi}Im\frac{W(\mathbf{k},\omega)} {W^2-Y^2-\phi^2},
 \nonumber \\ A_{X}(\mathbf{k},\omega)=
-\frac{1}{\pi}Im\frac{Y(\mathbf{k},\omega)} {W^2-Y^2-\phi^2},
 \nonumber \\ A_{\phi}(\mathbf{k},\omega)=
-\frac{1}{\pi}Im\frac{\phi(\mathbf{k},\omega)} {W^2-Y^2-\phi^2},
 \ea
where $\xi(\mathbf{k})$ is the bare dispersion,
 \ba
W(\mathbf{k},\omega)=\omega-\widetilde{\Sigma}(\mathbf{k},\omega)=\omega
Z(\vk,\omega) , \nonumber \\
Y(\mathbf{k},\omega)=\xi(\mathbf{k})+X(\mathbf{k},\omega),
 \ea
and $Z(\vk,\omega) $ is the renormalization function that
appears, for example, in the gap function
 \ba
\Delta(\vk,\omega)=\phi(\vk,\omega)/Z(\vk,\omega).
 \ea

\begin{figure}
\label{FS}
 \includegraphics[width=8cm]{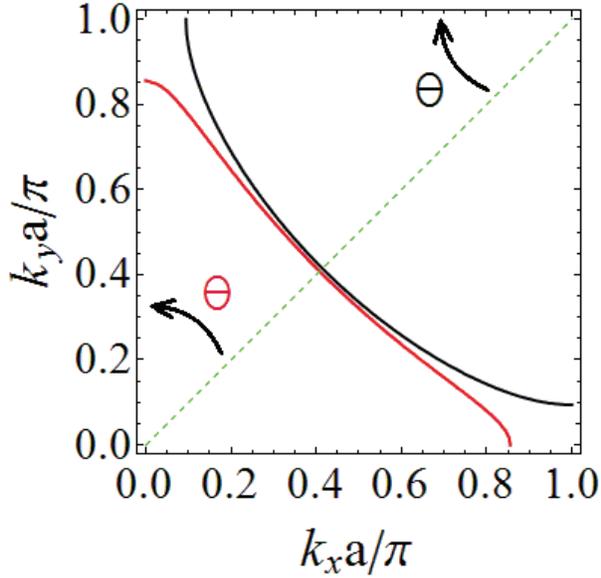}
\caption{The Fermi surface of OP LSCO (black line) and OV LSCO
(red line) given by Eq.\ (\ref{baredisp}). The OP and UD LSCO
have an hole-like and OV LSCO has a electron-like Fermi surface,
respectively, as can be seen from the figure. The tilt angle
$\theta$ is given with respect to the nodal cut centered at
$(\pi,\pi)$ for OP and UD, and at $(0,0)$ for OV LSCO. }
\end{figure}

The $4\times 4$ matrix self-energy may be written
as\cite{Maki69superconductivity}
 \ba
\hat{\Sigma}(\vk,\omega) = \widetilde{\Sigma}(\vk,\omega) \tau_0
+X(\vk,\omega) \tau_3 +\phi(\vk,\omega) \tau_2 \sigma_2 ,
 \ea
where $\tau_i$ and $\sigma_i$ are the Pauli matrices in the
particle-hole and spin space, respectively. The diagonal
self-energy is given by
 \ba
\Sigma(\mathbf{k},\omega) = \widetilde{\Sigma}(\mathbf{k},\omega)
+ X(\mathbf{k},\omega),
 \ea
and the diagonal spectral function measured by the ARPES is
 \ba
 \label{spectral}
A(\mathbf{k},\omega) = A_S(\mathbf{k},\omega)
+A_X(\mathbf{k},\omega).
 \ea
The symmetry of the self-energy is as follows.
 \ba
\widetilde{\Sigma}(\mathbf{k},\omega) &=&
-\widetilde{\Sigma}^*(\mathbf{k},-\omega), \nonumber \\
 X(\mathbf{k},\omega) &=& X^*(\mathbf{k},-\omega), \nonumber \\
\phi(\mathbf{k},\omega) &=& \phi^*(\mathbf{k},-\omega).
 \ea

We choose the following tight-binding band as the bare dispersion
for the doped \LSCO .\cite{Yoshida06prb}
 \ba
 \label{baredisp}
\xi(\vk) = -2t(\cos k_{x}a + \cos k_{y}a) + 4t'\cos k_{x}a \cos
k_{y}a \nonumber \\ - 2t''(\cos 2k_{x}a  + \cos 2k_{y}a ) - \mu.
 \ea
The tight-binding parameters are $t = 0.25$ eV, $t' =0.15~ t$,
$t'' =0.5~ t'$, and the chemical potential $\mu = -0.19$ eV for
OP, and $t = 0.25$ eV, $t' =0.13 ~t$, $t'' =0.5~ t'$, $\mu =
-0.22$ eV for OV, and $t = 0.25$ eV, $t' =0.17 ~t$, $t'' =0.5~
t'$, $\mu = -0.15$ eV for UD \LSCO . The UD and OP \LSCO \ have a
hole-like and OV has an electron-like FS as shown in Fig.\ 1. To
denote the momentum direction in the BZ in two dimensions we use
the tilt angle $\theta$ with respect to the nodal cut, that is,
the diagonal cut along $(0,0)-(\pi,\pi)$ line, centered at
$(\pi,\pi)$ for the hole like FS and at (0,0) for electron like
FS as indicated in Fig.\ 1.

The spin fluctuation mechanism in this formulation means that we
take $ F_{ch}({\bf q},\omega)=0$ and $F_{sp}({\bf q},\omega)$ as
the spin susceptibility $\chi_{sp}({\bf q},\omega)$ measured by
INS in Eq.\ (\ref{eliash-fun}). Then the imaginary parts of the
self-energy may be rewritten from Eq.\ (\ref{eliashberg}) as
%\begin{widetext}
 \ba
 \label{self-en2}
Im \widetilde{\Sigma}(\mathbf{k},\omega) = \pi
\int^{\infty}_{-\infty}d\epsilon'
\left[f(\omega-\epsilon')+n(-\epsilon') \right] \sum_{\mathbf{k'}}
A_{S}(\mathbf{k'},\omega-\epsilon')\alpha^{2}\chi_{sp}(\mathbf{k}-\mathbf{k'},\epsilon'),
 \nonumber \\
Im X(\mathbf{k},\omega)= \pi\int^{\infty}_{-\infty}d\epsilon'
\left[f(\omega-\epsilon')+n(-\epsilon') \right] \sum_{\mathbf{k'}}
A_{X}(\mathbf{k'},\omega-\epsilon')\alpha^{2}\chi_{sp}(\mathbf{k}-\mathbf{k'},\epsilon'),
 \nonumber \\
Im \phi(\mathbf{k},\omega)= \pi\int^{\infty}_{-\infty}d\epsilon'
\left[f(\omega-\epsilon')+n(-\epsilon') \right] \sum_{\mathbf{k'}}
A_{\phi}(\mathbf{k'},\omega-\epsilon')\alpha^{2}\chi_{sp}(\mathbf{k}-\mathbf{k'},\epsilon').
 \ea
%\end{widetext}
The real parts were calculated from the imaginary parts using the
Kramers-Kronig (KK) relation. Eqs.\ (\ref{self-en2}) and
(\ref{spectral-fun}) were solved self-consistently via iterations.

The coupling strength $\alpha$ was chosen such that it reproduces
the experimentally measured gap amplitude $\Delta_0$ of
\LSCO.\cite{Yoshida12jpsj} The coupling strength will be given in
terms of the dimensionless coupling constant $\lambda$ below,
 \ba
\lambda = \int_{-\infty}^\infty d\omega N(\omega) \frac{ \alpha^2
\chi_{sp}(\omega) }{\omega},
 \ea
where $N(\omega)$ is the density of states (DOS) and
$\chi_{sp}(\omega)$ is the local spin susceptibility given by
 \ba
\chi_{sp}(\omega)=\frac{\int d{\bf q} \ \chi_{sp}({\bf
q},\omega)} {\int d{\bf q}}.
 \ea
It is a measure of the density of spin excitation for a given
energy. The gap function is determined by
 \ba
\Delta(\vk) = Re \left[\frac{\phi(\vk,\omega)}{Z(\vk,\omega)}
\right]_{\omega=\Delta(\vk)},
 \ea
and $\Delta_0$ was determined from the DOS peak position.

The measured spin susceptibility $\chi_{sp}({\bf q},\omega)$ from
INS was fitted by the form
 \ba
\chi_{sp}({\bf q},\omega) = \chi_\delta (\omega)
\frac{\kappa^4(\omega)}
{\left[\kappa^2(\omega) +R({\bf q}) \right]^2}, \nonumber \\
 R({\bf q}) = \frac{\left[ (h-{\tfrac12})^2 +(k-\tfrac12)^2-\delta^2 \right]^2
+\lambda(h-\tfrac12)^2(k-\tfrac12)^2  } {4\delta^2},
 \ea
where the inplane wave-vector is written in the reciprocal lattice
as ${\bf q} = h{\bf a}^* +k{\bf b}^* +l{\bf c}^* $,
$\kappa(\omega)=1/\xi$ the inverse correlation length, the
incommensurability $\delta(\omega)$ specifies the position of the
four peaks, and $\lambda(\omega)$ controls the shape of the
pattern. $\lambda$ = 4 corresponds to four distinct peaks and
$\lambda$ = 0 corresponds to a pattern with circular symmetry.
These fitting parameters were given in the references
\cite{Vignolle07naturephys,Lipscombe07prl,Lipscombe09prl,Lipscombe08thesis}.
Notice, however, that in these references the authors used the
wave-vectors in the reciprocal lattice unit such that their 1/2,
for example, corresponds to $\pi$ of this paper.

The measured local susceptibility may be decomposed into three
parts; a low frequency incommensurate (IC) peak centered around
${\bf Q}_\delta =(\pi\pm\delta,\pi)$ and the symmetry related
points, a commensurate (CM) peak at ${\bf Q}=(\pm\pi,\pm\pi)$,
and a broad high frequency feature. The IC peak is around 18, 15,
and 15 meV for OP, OV, and UD samples, respectively. The CM peak
is around 50 meV for OP and 45 meV for UD samples, but is missing
for OV \LSCO. On the other hand, the high frequency feature
persisting up to measurable energy is common for all samples. The
cutoff energy $\omega_c$ of the susceptibility spectrum was taken
to be 0.3 eV. This is the upper limit of the spin wave
spectrum\cite{Headings10prl} of around $~2J$.

The $\vk'$ summation in Eq.\ (\ref{self-en2}) was performed by
using the 2D fast Fourier transform (FFT) between the momentum and
real space using the convolution relation
 \ba
\sum_{\vk}e^{i \vk\cdot{\bf r}} \sum_{\vk'} F(\vk'-\vk) G(\vk') =
F({\bf r}) G({\bf r})
 \ea
on a $2^{8}\times2^{8}$ mesh of the first quadrant of BZ. No
assumption about the $\vk$ and $\omega$ dependence nor a
separable form of the diagonal and off-diagonal self-energy was
made in the calculations. Self-consistency is reached in a couple
of tens of iterations.

\section{Preliminary Analysis}

Before presenting our results of the angle and frequency
dependence of the self-energy, it will be useful to consider some
simple cases. Let us first consider the Einstein model of
frequency $\omega_b$ of the coupled boson.
 \ba
\alpha^2 F(\vk,\vk',\epsilon')= \alpha^2(\vk,\vk')
 \left[ \delta(\epsilon'-\omega_b)- \delta(\epsilon'+\omega_b)
\right].
 \ea
Then the imaginary part of the diagonal self-energy from Eq.\
(\ref{self-en2}) is
 \ba
 \label{self2}
\Sigma_2(\vk,\omega) = \pi \left\{ \left[ f(\omega-\omega_b)
+n(-\omega_b) \right] D(\vk,\omega-\omega_b) \right. \nonumber \\
\left.
 -\left[
f(\omega+\omega_b) +n(\omega_b) \right] D(\vk,\omega+\omega_b)
\right\},
 \ea
where
 \ba
 \label{weighteddos}
D(\vk,\omega)=\sum_{\vk'} \alpha^2(\vk,\vk') A(\vk',\omega).
 \ea
In the low temperature limit of $T\rightarrow 0$, it is reduced to
 \ba
 \label{self2T0}
-\Sigma_2(\vk,\omega)= \pi \left[\Theta(\omega -\omega_b)
D(\vk,\omega-\omega_b) + \Theta(-\omega -\omega_b)
D(\vk,\omega+\omega_b) \right],
 \ea
where $\Theta$ is the step function. The peaks of $-\Sigma_2
(\vk,\omega)$ for the negative (positive) $\omega$ region are
determined by those of $D(\vk,\omega\pm \omega_b)$. Depending on
the range of $\vk'$ summation  of Eq.\ (\ref{weighteddos})
determined by $\alpha^2(\vk,\vk')$, either one peak (for $\kappa
\rightarrow \infty$ or 0) or two peaks (for intermediate
$\kappa$) may show up as discussed below.

Consider two limits of this expression: First, for momentum
independent coupling of $\alpha^2(\vk,\vk') = \alpha^2 $. Then,
we have
 \ba
 \label{self2dos}
-\Sigma_2(\vk,\omega)= \left\{
 \begin{array}{ll}
\pi \alpha^2 N(\omega-\omega_b), & {\rm for~} \omega>\omega_b, \\
\pi \alpha^2 N(\omega+\omega_b), & {\rm for~} \omega< - \omega_b,
\\
 0, & {\rm otherwise}.
 \end{array} \right.
 \ea
where
 \ba
N(\omega)=\sum_{\vk'} A(\vk',\omega)
 \ea
is DOS. This clearly shows that the peaks of DOS at $\omega=\pm
\Delta_0$ in the SC state are shifted to $\pm(\Delta_0
+\omega_b)$ in $-\Sigma_2(\omega)$ because of the coupling to the
boson of frequency of $\omega_b$ and that they are momentum
independent. This case is relevant where the correlation length
$\xi$ of a susceptibility peak is small (or, the inverse
correlation length $\kappa \rightarrow \infty$) like the 50 meV CM
peak of OP sample. See the angle independent peaks near $|\omega|
\approx 65$ meV in $-\Sigma_2$ and $\phi_2$ shown in Figs.\ 2(a)
and 3(a).

Another limit is where the coupling is a delta function like
$\alpha^2(\vk,\vk') = \alpha^2 ~ \delta(\vk'-\vk-{\bf q})$,
corresponding to $\kappa \rightarrow 0$. Then, instead of Eq.\
(\ref{self2dos}), we have
 \ba
 \label{qdelta}
-\Sigma_2(\vk,\omega)= \left\{
 \begin{array}{ll}
\pi \alpha^2 A(\vk+{\bf q},\omega-\omega_b), & {\rm for~} \omega>\omega_b, \\
\pi \alpha^2 A(\vk+{\bf q},\omega+\omega_b), & {\rm for~} \omega<
- \omega_b, \\
 0, & {\rm otherwise}.
 \end{array} \right.
 \ea
Because the spectral function $A(\vk,\omega)$ has a peak around
$\omega\approx \pm E(\vk)$ in SC state, where
 \ba
E(\vk)=\sqrt{\widetilde{\xi}^2(\vk) +\Delta^2(\vk)},
 \ea
and $\widetilde{\xi}(\vk)=(\xi(\vk)+X)/Z$ is the renormalized
dispersion, the peak of $-\Sigma_2(\vk,\omega)$ occurs at
$\omega\approx \pm (E(\vk+{\bf q})+\omega_b )$ which is clearly
momentum and band structure dependent.

For intermediate values of $\kappa$, the self-energy of Eq.\
(\ref{qdelta}) is summed around $\vk+{\bf q}$ over the width
$~\kappa$. Then, both energy scales of $(\Delta_0 +\omega_b)$ and
$(E(\vk+{\bf q})+\omega_b )$ may appear in
$-\Sigma_2(\vk,\omega)$ from a single peak of $\chi({\bf
q},\omega)$. A modification is that the energy $(\Delta_0
+\omega_b)$ of $\xi\rightarrow 0$ now becomes momentum dependent
$(\Delta'(\vk) +\omega_b)$ because a non-zero $\xi$ implies a
momentum selection in the $\vk'$ summation. $\Delta'(\vk)$ is an
angle dependent energy of order $\Delta_0$. This seems to be the
case for the IC peaks as will be discussed below.

Some complications arise from the band structure and momentum
dependent coupling.\cite{Sandvik04prb} An interesting case is
where the momentum sum covers the saddle point which usually
occurs at $(0,\pi)$. This introduces another energy scale in
$-\Sigma_2(\vk,\omega)$ from the van Hove singularity (VHS). It
occurs at $\omega\approx sgn(\xi_{VHS})(E_{VHS} +\omega_b)$, where
$E_{VHS}=\sqrt{\widetilde{\xi}^2(0,\pi)+\Delta_0^2}$ and
$sgn(f)=\pm 1$ represents the sign of $f$, in addition to the two
energy scales of the peaks discussed above. The shape of
$\Sigma(\vk,\omega)$ are also modified by the impurity
scatterings.\cite{Zhu04prb} The VHS peak may be substantially
suppressed by the coupling to boson spectrum and impurity
scatterings. One should perform the self-consistent calculations
to see their effects without misleading conclusion. The off-plane
elastic impurities may induce interesting features of the
self-energy in the SC state.

In cases where the VHS peak is strongly suppressed and/or the
$E_{VHS}$ and $\Delta_0$ are not well separated, the VHS feature
may not clearly show up. The parameters of current LSCO
calculations seem to belong to this case and we do not discuss
the VHS features in the self-energy below. Also recall that the
discussion so far is restricted to a sharp boson frequency of a
single energy. A finite width in energy as well as in momentum
space smoothens peak features in the self-energy.

\section{Numerical Results }

We now turn to the self-consistent numerical calculations using
the experimentally measured spin susceptibility as the Eliahberg
function for the OP, OV, and UD \LSCO \ as explained above. The
angle $\theta$ in the BZ was chosen with respect to the nodal
direction as shown in Fig.\ 1. We wish to discuss the position
and intensity of peaks in the absolute value of the imaginary
part, $-\Sigma_2(\vk,\omega)$, and the real part of the
self-energy, $\Sigma_1(\vk,\omega)$, for $\omega<0$. The
$\omega>0$ region can not be probed by the ARPES with which we
wish to compare our numerical results. For the off-diagonal
self-energy, $\phi_1(\vk,\omega)$ and $\phi_2(\vk,\omega)$ are,
respectively, even and odd functions of $\omega$, and will be
shown in the $\omega>0$ region.

\subsection{OP \LSCO }

First, we consider the OP \LSCO\ with the doping concentration
$x=0.16$ and the critical temperature $ T_c =38.5$ K. The spin
susceptibility spectrum reported by Vignolle $et~al.$ has three
parts; the IC peak near 18 meV, CM peak near 50 meV, and broad
high frequency feature extended to 0.3 eV. The coupling constant
was chosen such that $\lambda=1.74$ in the calculations to obtain
the gap amplitude $\Delta_0=17$ meV in the $T\rightarrow 0$
limit.\cite{Yoshida12jpsj}

\begin{figure}[tbh]
\includegraphics[width=8cm]{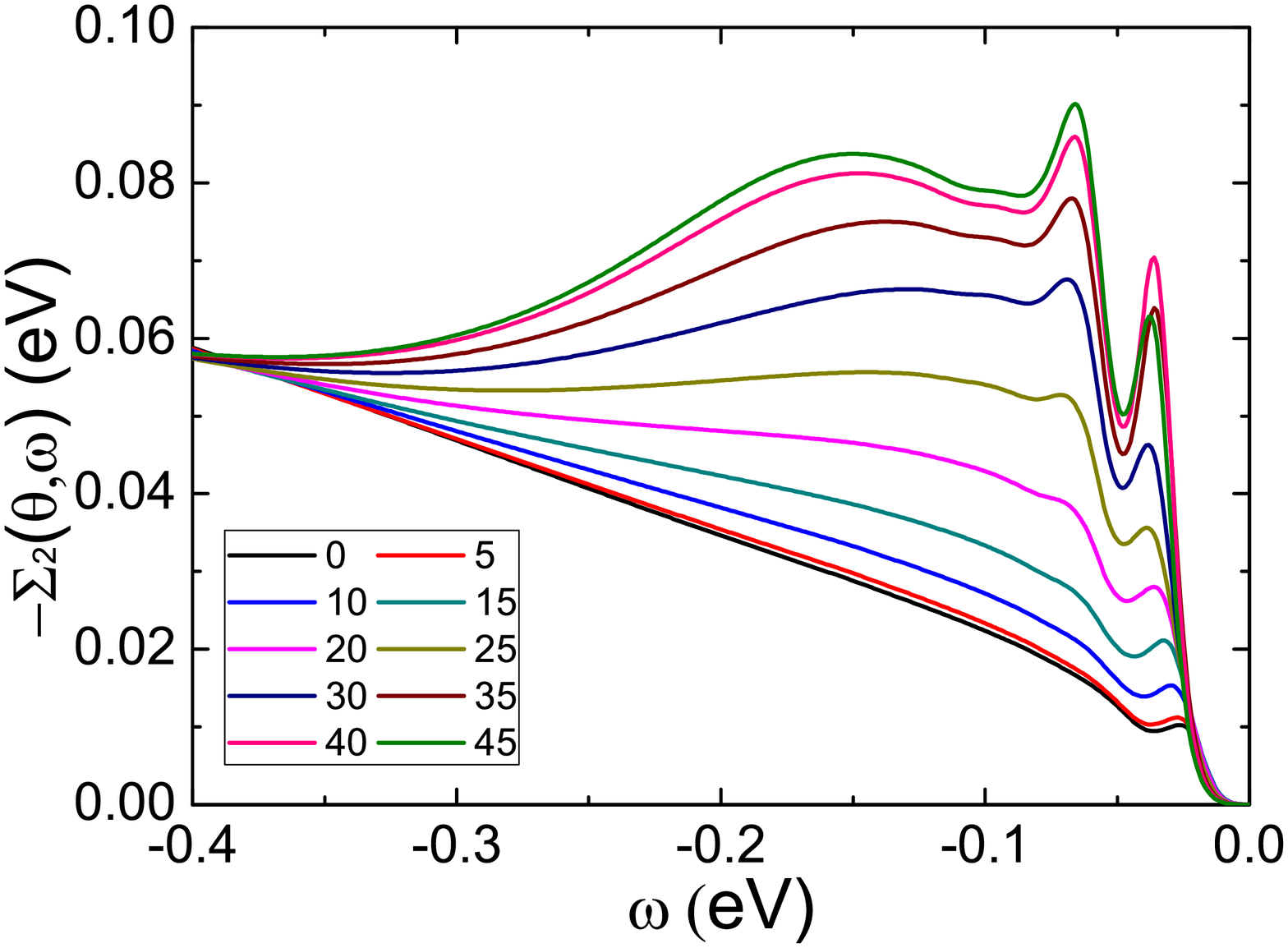}
\includegraphics[width=8cm]{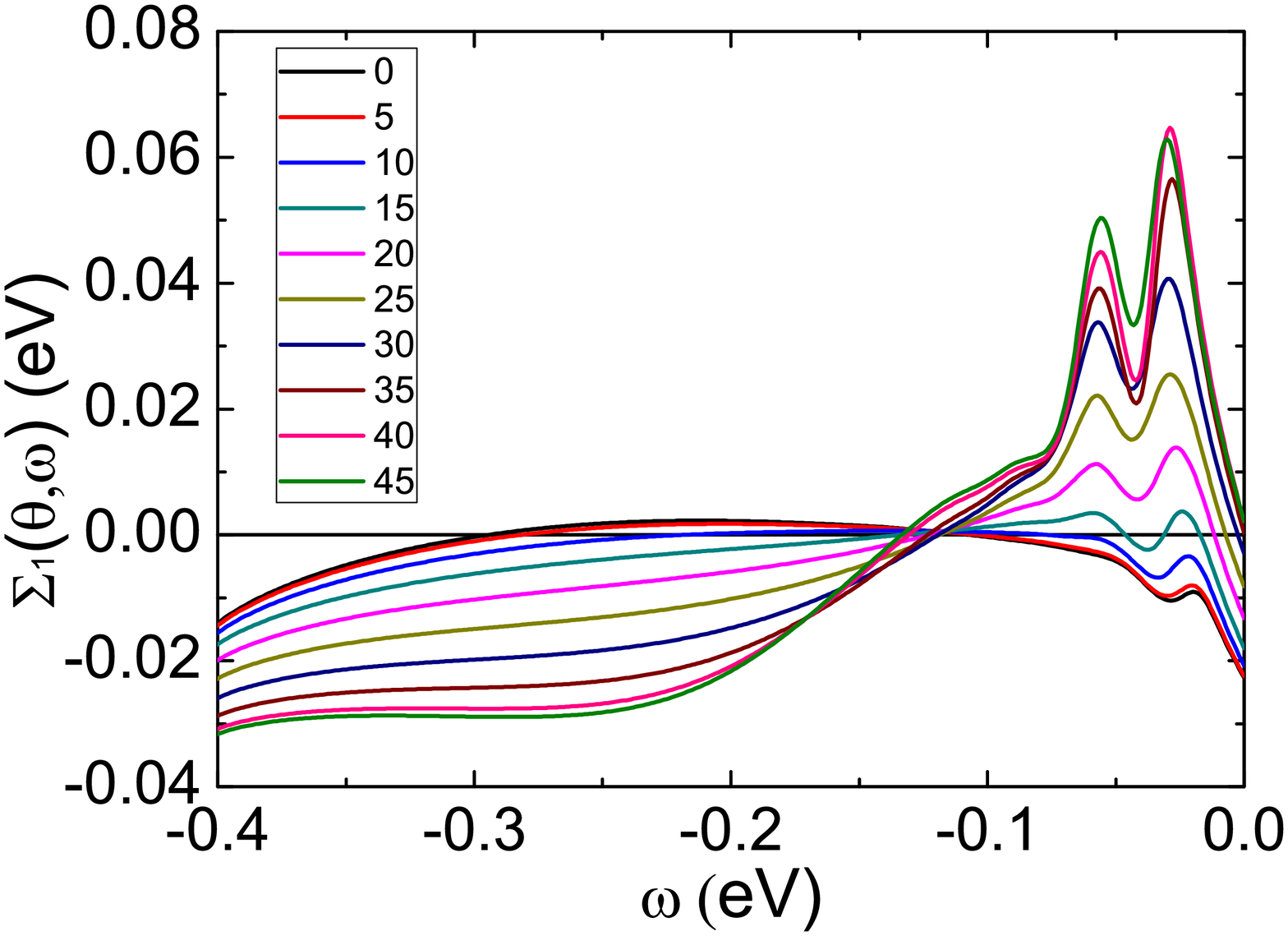}
\caption{The calculated imaginary and real parts of the diagonal
self-energy at the Fermi momentum of OP LSCO along several cuts
in BZ in the SC state. The Vignolle spectrum was used for the
$F_{sp}$ in the Eliashberg equation. The nodal direction is 0 deg
and the anti-nodal direction is 45 deg.}
\end{figure}

Fig.\ 2(a) and (b) are the imaginary and real parts of the
diagonal self-energy along several cuts perpendicular to the FS
in the BZ in SC state. Fig.\ 3(a) and (b) show the imaginary and
real parts of the off-diagonal self-energy, respectively. The two
peaks in both the real and imaginary parts of the diagonal and
off-diagonal self-energies reflect the two peaks in the spin
susceptibility with slight complication as explained below.

\begin{figure}[tbh]
\includegraphics[width=8cm]{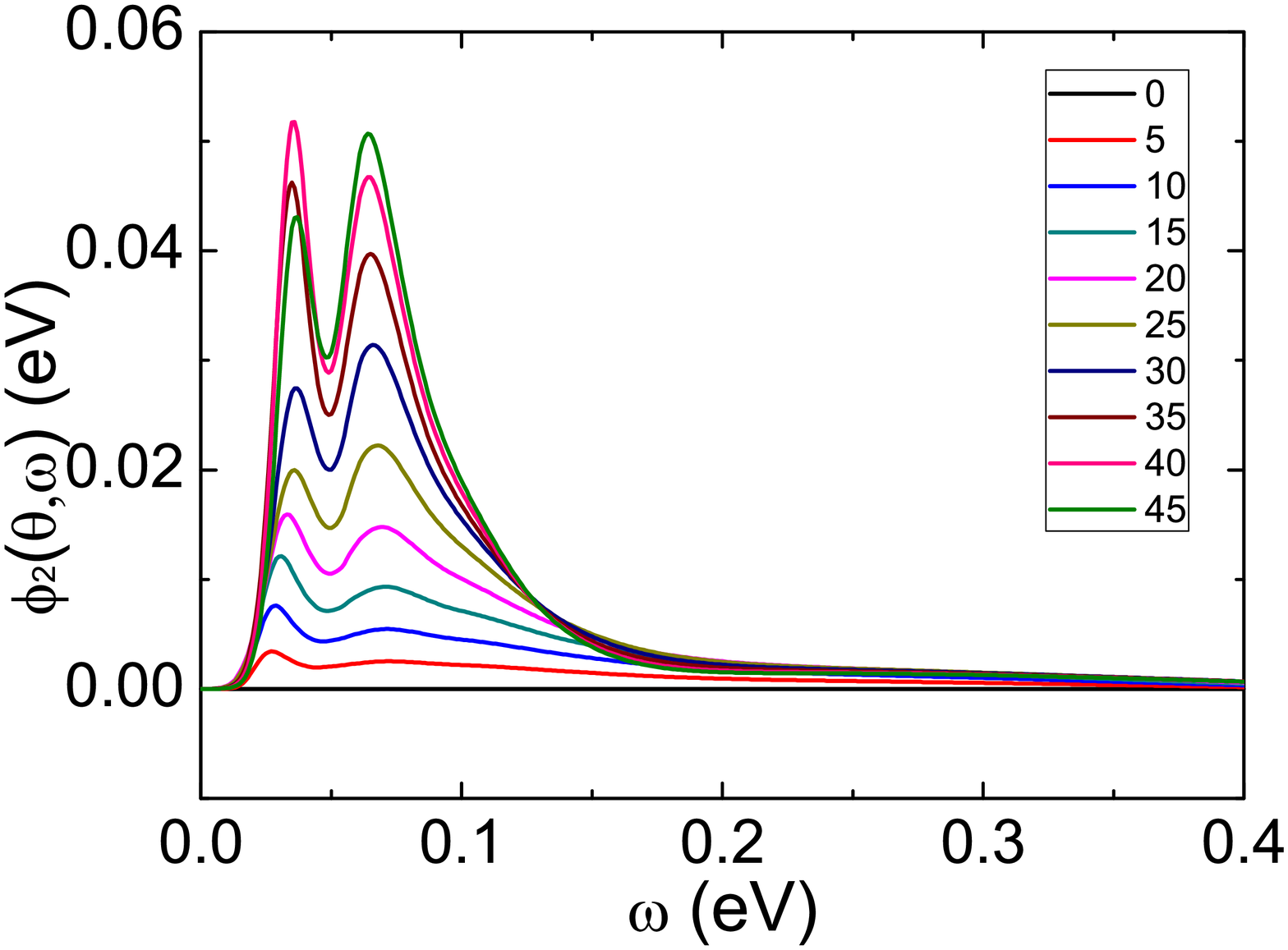}
\includegraphics[width=8cm]{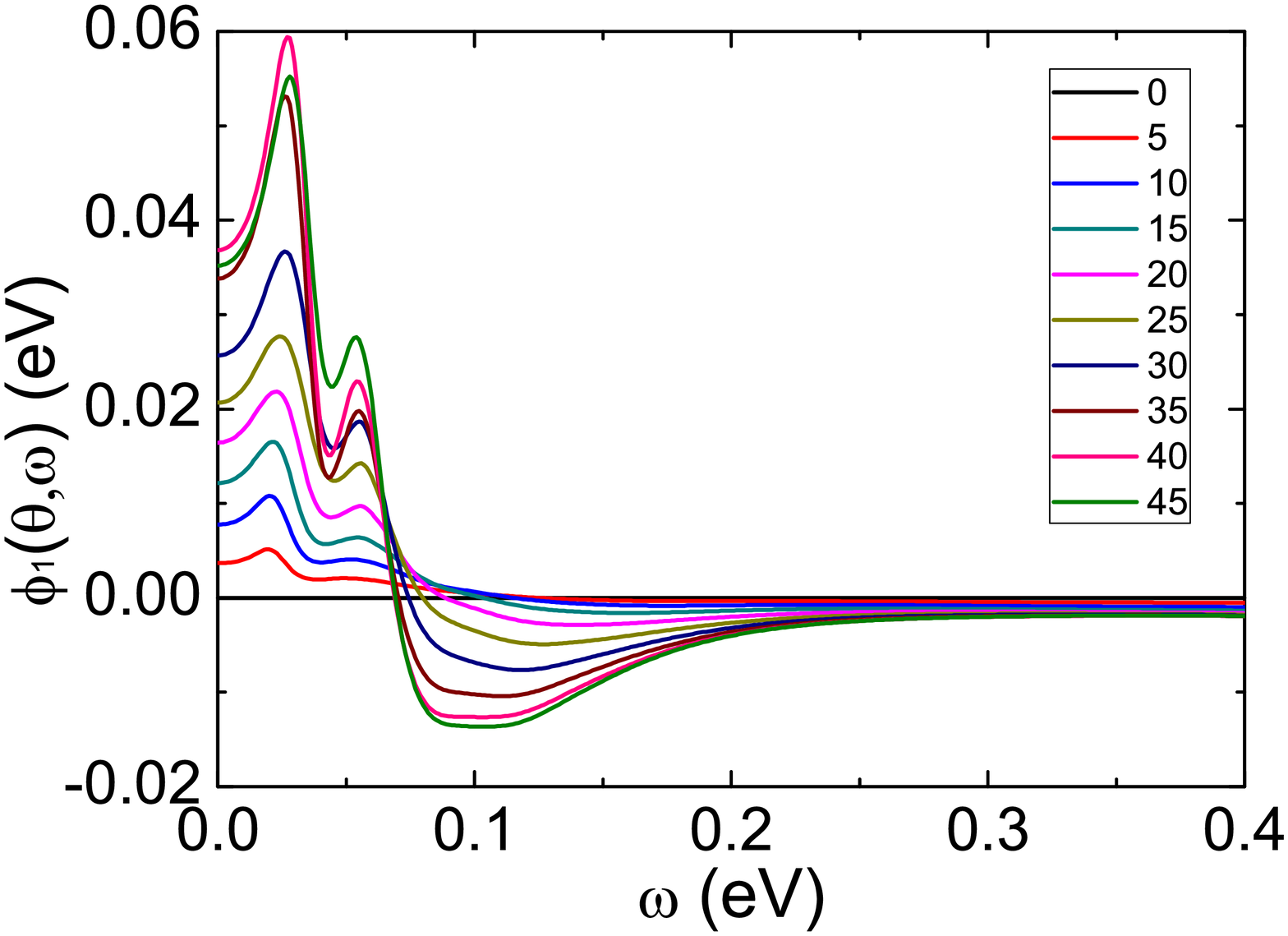}
\caption{The calculated imaginary and real parts of the
off-diagoanl self-energy of OP LSCO along several cuts
perpedicular to the FS in the SC state. The angles are the same as
the Fig.\ 2.}
\end{figure}

For the Vignolle spectrum from OP \LSCO, the IC peak has an
intermediate correlation length of $\xi\approx 4.1a$, the CM peak
has a small $\xi \approx 0.8 a$, and the broad high frequency
feature has $\xi \approx 0.7 a$.\cite{Vignolle07naturephys} From
the discussion in the previous section, we may expect two peaks at
$\omega_1 \approx -(\Delta'(\vk) +\omega_{IC})$ and
$\omega'_1\approx -(E(\vk+{\bf Q_\delta}) +\omega_{IC})$ from the
IC peak, and one peak at $\omega_2 \approx -(\Delta_0
+\omega_{CM})$ from CM peak. The $\omega'_1$ seems to overlap
with $\omega_2$, and two peaks show up in
$-\Sigma_2(\vk,\omega)$. The $\omega_1$ is expected to be
angle-dependent and $\omega_2$ to be $\approx -67$ meV because
$\Delta_0 =17$ and $\omega_{CM}\approx 50$ meV. This is what we
obtained in numerical calculations as shown in Fig.\ 2(a). The
exactly same argument holds for the off-diagonal self-energy,
$\phi_2 (\vk,\omega)$ as shown in Fig.\ 3(a).

In order to understand the peak energy of the real parts of the
self-energy, recall that the real and imaginary parts are related
by the KK relation. It means that the peak energy of
$\Sigma_1(\omega)$ is shifted from that of $-\Sigma_2(\omega)$ by
the width of the peak, that is, the peak energy of $\Sigma_1$ is
expected at $\omega\approx -(\Delta'(\vk)
+\omega_{IC}-\Gamma_{IC})$ and $\omega\approx -(\Delta_0
+\omega_{CM}-\Gamma_{CM})$, where $\Gamma$ is the width of the
peak. This is indeed what we obtained from numerical calculations.
See the plots of $\Sigma_1(\vk,\omega)$ and $\phi_1(\vk,\omega)$
as shown in Figs.\ 2(b) and 3(b), respectively.

Now, we turn to the intensity of the peaks of the self-energy.
From Eqs.\ (\ref{self2T0}) and (\ref{weighteddos}) we see that
$-\Sigma_2(\vk,\omega)$ is given by the sum over $\vk'$ of
$\alpha^2(\vk,\vk') A(\vk',\omega+\omega_b)$. It means that there
is large contribution from the $\vk'$ sum to
$-\Sigma_2(\vk,\omega)$ if both $\vk'-\vk\approx {\bf Q}_\delta$
and $\omega \approx -(E(\vk+{\bf Q}_\delta)+\omega_{IC})$ are
satisfied. This is better satisfied near the anti-nodal region and
$-\Sigma_2(\vk,\omega)$ increases as the tilt angle increases for
small $|\omega|$. For large $|\omega|$, however, either of the two
conditions become ill satisfied and $-\Sigma_2(\vk,\omega)$ is
roughly angle independent. This is indeed what Fig.\ 2(a) shows.
The IC peak in all four plots in Fig.\ 2 and Fig.\ 3 is not
highest at 45 deg but $\approx 40$ deg because of the
incommensurability $\delta$. On the other hand, the CM peak near
65 meV is highest at 45 deg as expected because the broad CM peak
connects the anti-nodal regions most effectively.

The angle dependence of the off-diagonal self-energy
$\phi(\theta,\omega)$ along several cuts is roughly $d$-wave like
as shown in Fig.\ 3. The imaginary part looks like the local spin
susceptibility $\chi_{sp}(\omega)$ with the suppressed high
frequency part. The suppression of $\phi_2(\theta,\omega)$ above
$\sim 0.1$ eV shows that the high frequency part of the
susceptibility does not contribute much to pairing because its
broad momentum dependence is not very effective for $d$-wave
pairing. The real part $\phi_1(\theta,\omega)$ increases as
$\omega$ increases from 0 and exhibits two peaks induced by the
two peaks in the spin susceptibility and then decreases and makes
a zero crossing near $\omega_2$. There, $\phi_2(\theta,\omega)$
has a peak because of the KK relation.

\subsection{OV \LSCO }

\begin{figure}[tbh]
\includegraphics[width=8cm]{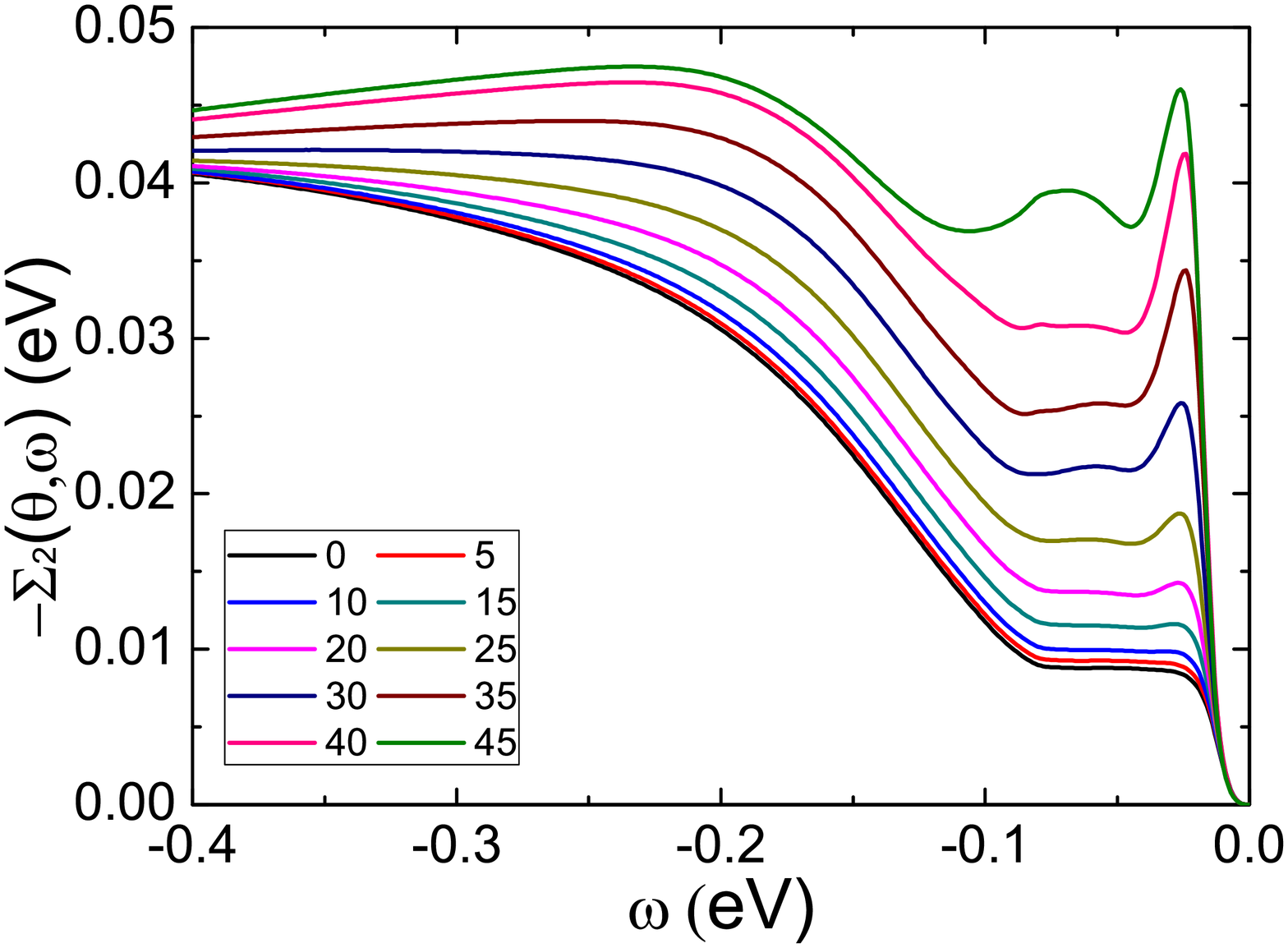}
\includegraphics[width=8cm]{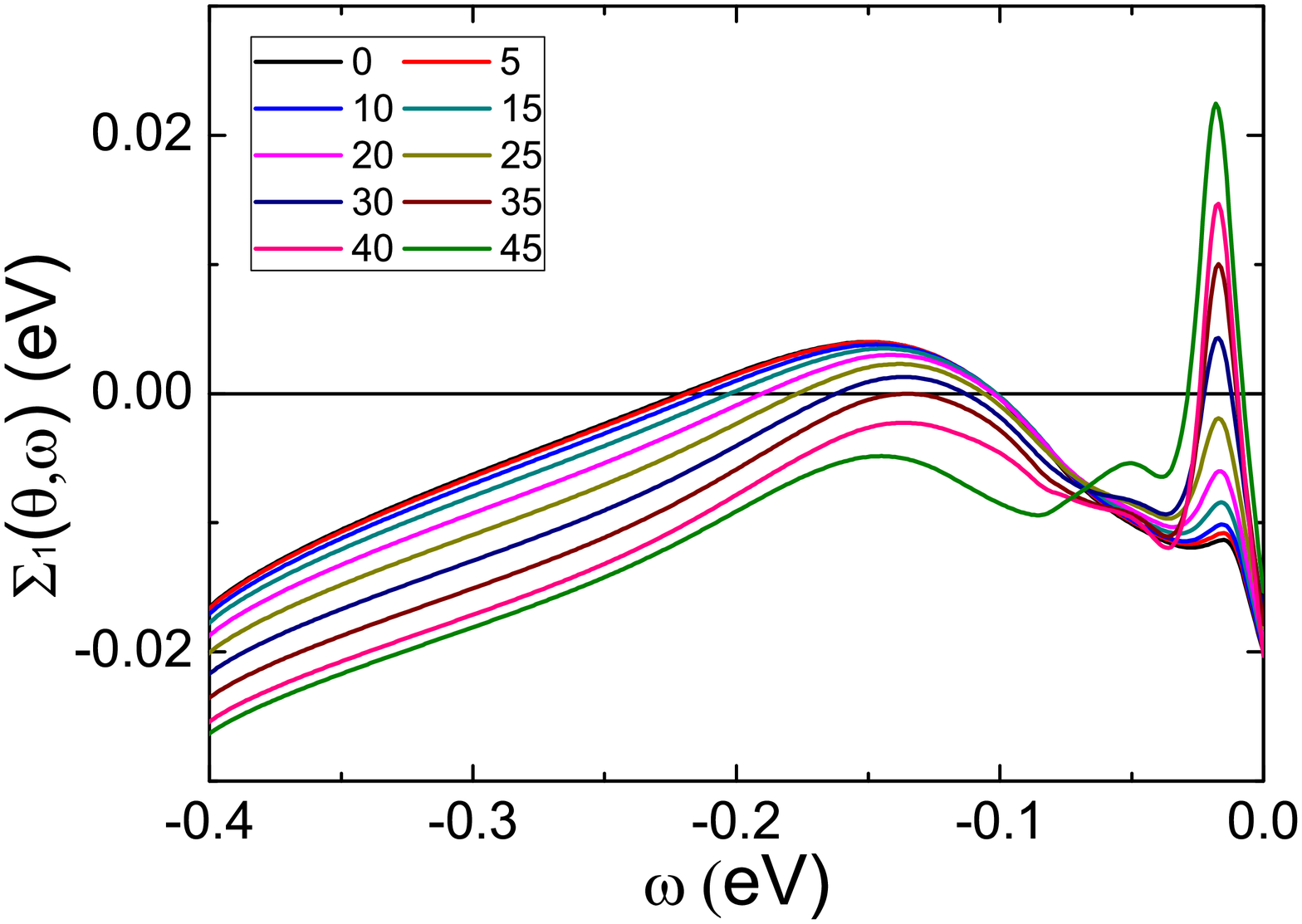}
\caption{The calculated imaginary and real parts of the diagonal
self-energy of OV LSCO along several cuts in BZ in the SC state. }
\end{figure}

The calculations were done for OV \LSCO\ as well. The difference
from the OP case is that (a) the spin susceptibility spectrum
does not have the CM component, and (b) the bare dispersion has
smaller next nearest neighbor hopping amplitude of $t'/t = 0.13$
and has an electron-like FS. The IC peak at 15 meV of the spin
suscetibility has the correlation length of $\xi \approx 2.7 a$
and the broad high energy feature has $\xi\approx 0.6
a$.\cite{Lipscombe09prl} The coupling constant was chosen such
that $\lambda=1.98$ to obtain the gap amplitude $\Delta_0=11$ meV
at $T=0$ in calculations.

\begin{figure}[tbh]
\includegraphics[width=8cm]{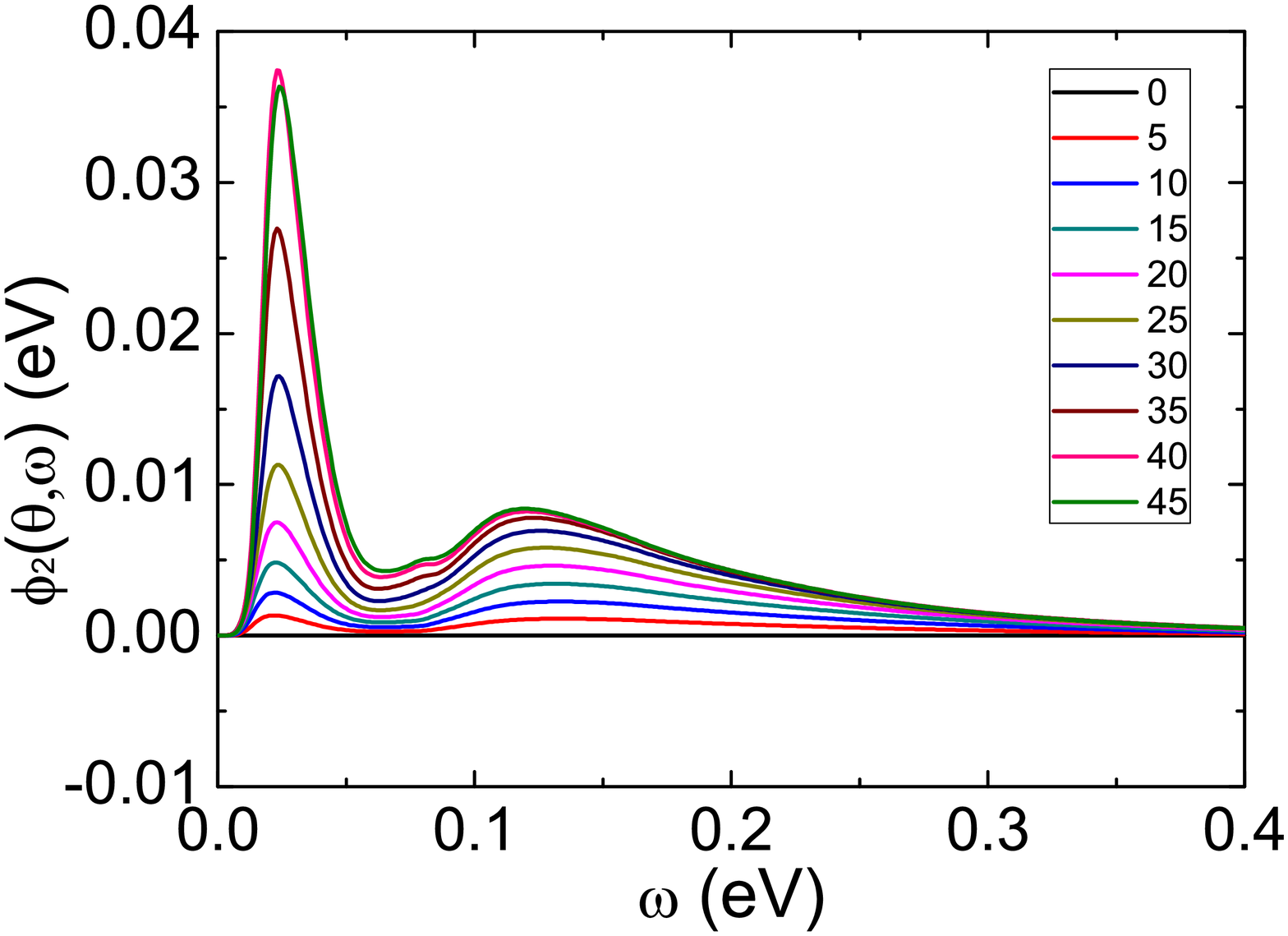}
\includegraphics[width=8cm]{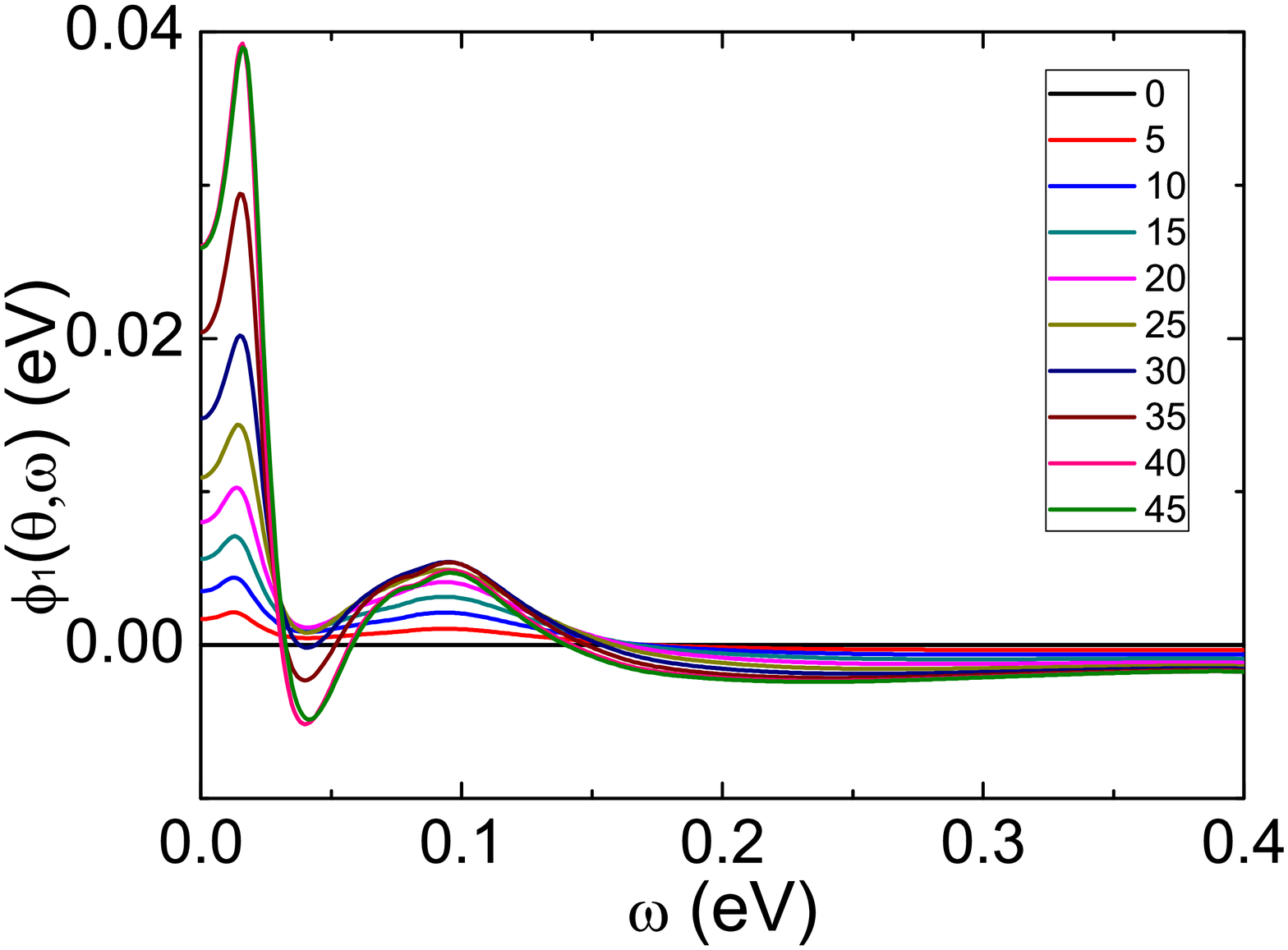}
\caption{ The calculated imaginary and real parts of the
off-diagoanl self-energy at the Fermi momentum of OV LSCO in the
SC state. }
\end{figure}

Fig.\ 4(a) and (b) are the imaginary and real parts of the
diagonal self-energy along several cuts perpendicular to the
Fermi surface in the BZ. The results may be understood like the
OP case. The single IC peak in the spin susceptibility has an
intermediate correlation length and induces two peaks in
$-\Sigma_2(\vk,\omega)$. Because the correlation length is rather
small, the angle dependence of the peak at $\omega_1 \approx -25$
and $\omega_2 \approx -75 $ meV are weak. The peak at $\omega_1$
is the shift of the DOS peak because $\Delta_0 +\omega_{IC}
\approx 26$ meV. The peak at $\omega_2$ is from the $-(E(\vk+{\bf
Q}_\delta)+\omega_{IC})$ as was discussed in the OP case. We note
that the VHS peak can not appear in the negative energy because
OV LSCO has electron-like Fermi surface.

Fig.\ 5(a) and (b) show the imaginary and real parts of the
off-diagonal self-energy along several cuts in BZ. The angle
dependence is roughly $d$-wave like. The imaginary part of the
off-diagonal self-energy $\phi_2(\theta,\omega)$ has a peak near
$\omega_1\approx 25$ meV and looks like the local spin
susceptibility with the suppressed high frequency part as the OP
case. The real part $\phi_1(\theta,\omega)$ begins to increase as
$\omega$ increases from 0, has a peak and then decreases and
makes a zero-crossing near $\omega_1$ where the
$\phi_2(\theta,\omega)$ has the IC peak.

\subsection{UD \LSCO }

The calculations were done for UD \LSCO\ as well. The spin
susceptibility spectrum reported by Lipscombe
$et~al.$\cite{Lipscombe09prl,Lipscombe08thesis} for 8\% doping
\LSCO\ was used in the calculations. The spectrum
$\chi_{sp}(\omega)$ has three parts as the OP case; the IC peak
near 15 meV, CM peak near 45 meV, and a broad high frequency
feature extrapolated upto 0.38 eV. The IC and CM peaks have the
correlation lengths of $\xi\approx 2.45a$ and $0.8a$,
respectively.\cite{Lipscombe08thesis} The coupling constant was
chosen such that $\lambda=1.58$ to obtain the gap amplitude
$\Delta_0=17$ meV at $T=0$ in the
calculations.\cite{Yoshida12jpsj}

\begin{figure}[tbh]
\includegraphics[width=8cm]{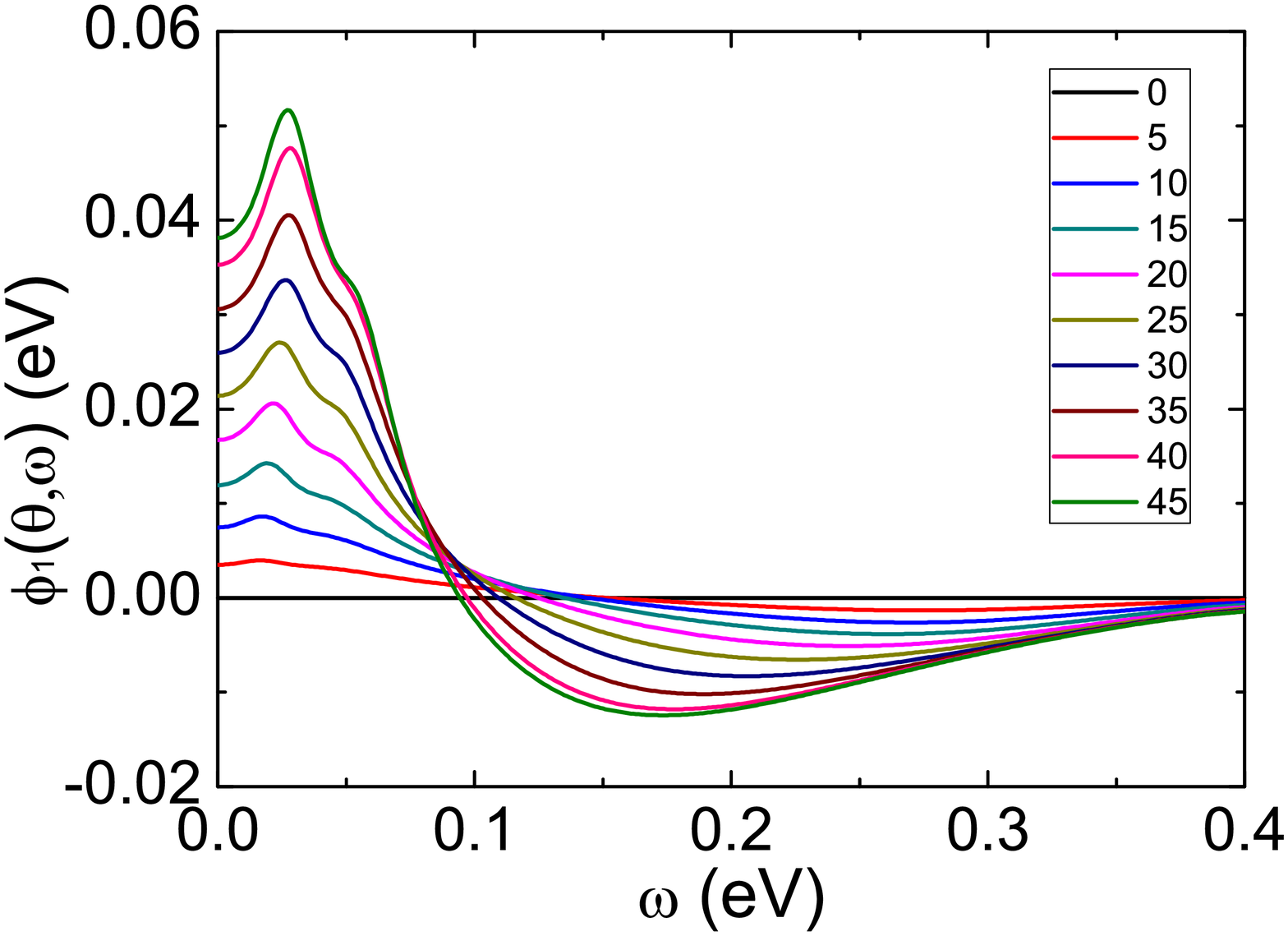}
\includegraphics[width=8cm]{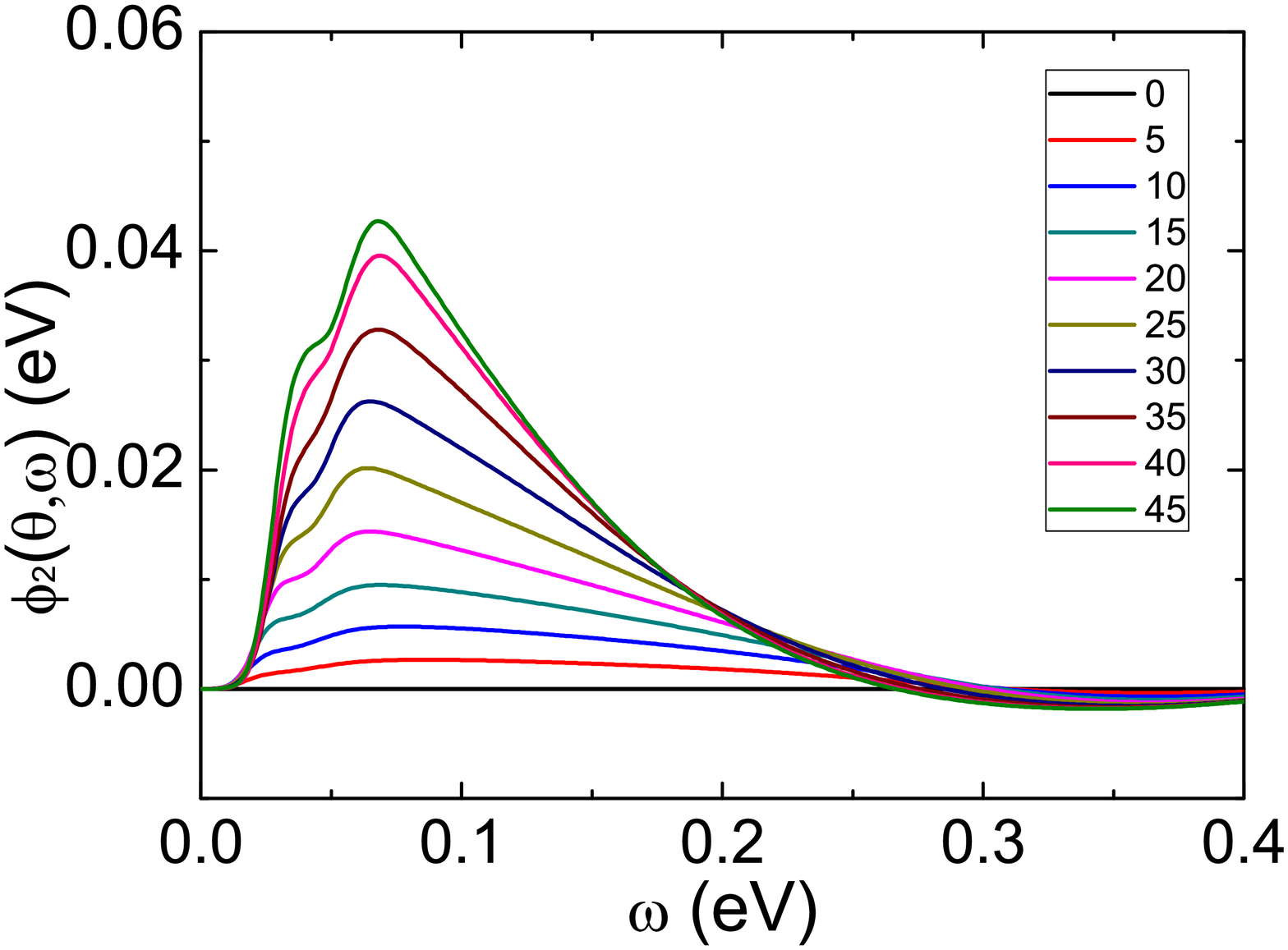}
\caption{(a), (b) The calculated imaginary and real parts of the
diagonal self-energy of UD LSCO along several cuts in BZ in the SC
state. }
\end{figure}

Fig.\ 6(a) and (b) are the imaginary and real parts of the
diagonal self-energy along several cuts perpendicular to the
Fermi surface in the BZ. The results may be understood like the
OP and OV cases. Because the peaks of the susceptibility are
broader in frequency than OP and OV materials, the peaks in the
self-energy are not as sharp as the OP and OV cases.

\begin{figure}[tbh]
\includegraphics[width=8cm]{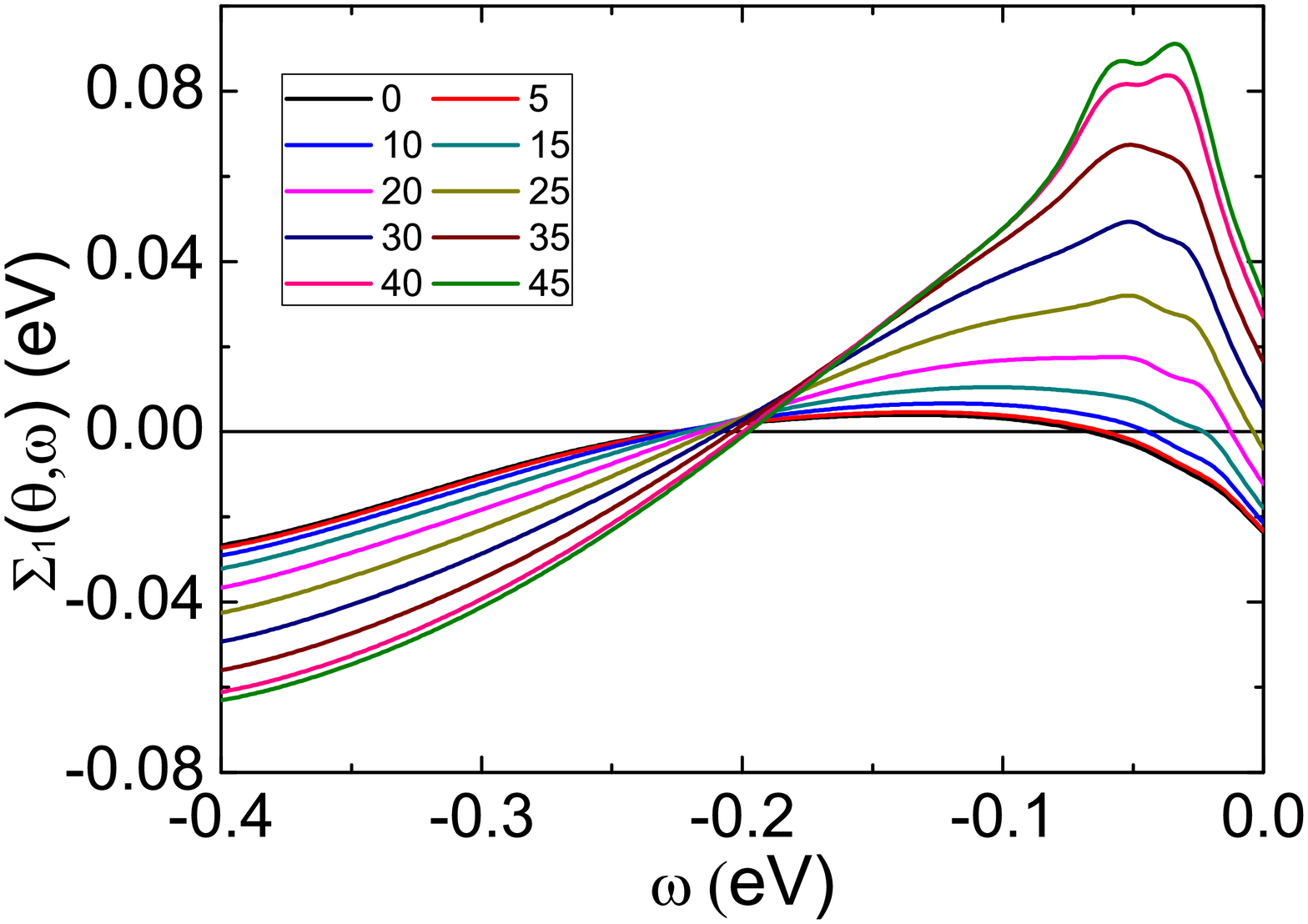}
\includegraphics[width=8cm]{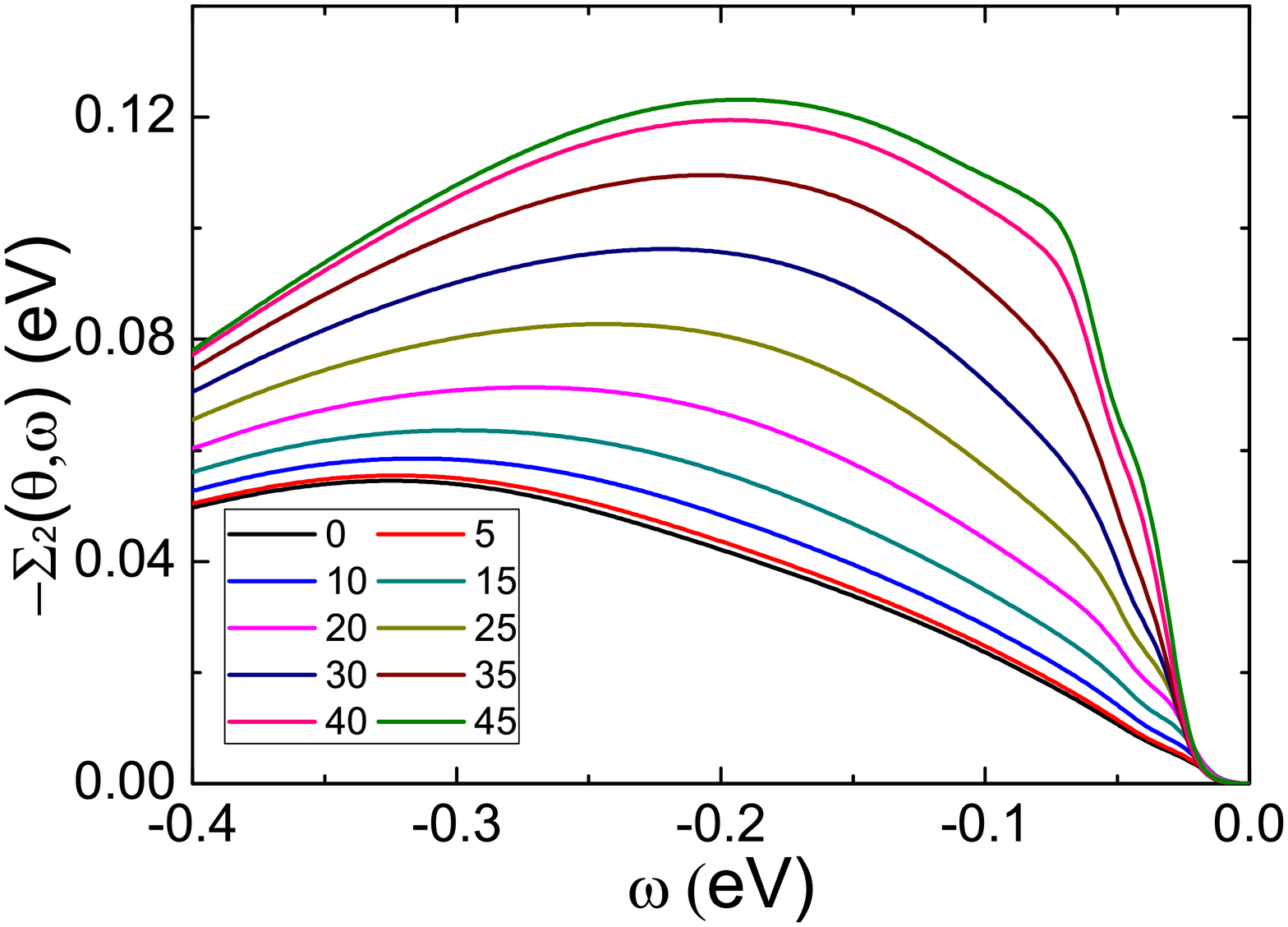}
\caption{(a), (b) The calculated imaginary and real parts of the
off-diagoanl self-energy. }
\end{figure}

Fig.\ 7(a) and (b) show the imaginary and real parts of the
off-diagonal self-energy along several cuts. The angle dependence
is roughly $d$-wave like. The two peaks in the diagonal and
off-diagonal self-energy are from the IC and CM peaks of the spin
susceptibility. Their energy in the imaginary parts is expected
at $\omega_1 \approx \Delta_0 +\omega_{IC} \approx 32$ and
$\omega_2 \approx \Delta_0 +\omega_{CM} \approx 62$ meV. The
peaks in the real parts are shifted by the width. This is what we
obtained from the calculations as shown in Figs.\ 6 and 7.

\section{Summary~ and ~Concluding~Remarks}\label{sec:conclusion}

We have calculated the full momentum and frequency dependence of
the self-energy by solving the Eliashberg equation using the
measured spin susceptibility from inelastic neutron scattering
experiments on optimally, overdoped, and underdoped \LSCO \
cuprates in the SC state. The real and imaginary parts of the
diagonal and off-diagonal self-energy were presented for several
cuts perpendicular to the Fermi surface for each doping
concentration.

The results of calculations were discussed in terms of the angle
(i.e. the direction of momentum in the BZ) dependence of the peak
position and intensity. First, the angle dependence of the peak
intensity is that the spin fluctuation induced self-energy
$\Sigma(\vk,\omega)$ is very anisotropic in the momentum space
for small $|\omega|$. We can see from Fig.\ 2(a) that the
absolute value of the imaginary part of the diagonal self-energy
of OP LSCO increases by a factor of about 5 from nodal to
anti-nodal directions below the CM peak of $|\omega| \leq 70$
meV. The real part also changes roughly by the same factor as can
be seen from the inset of Fig.\ 2(b). This is understandable
because the spin susceptibility with the correlation length $\xi
\approx 4.1 a$ must mean that the quasi-particle dynamics is very
different for different directions in the BZ. Second, The angle
dependence of the peak position is that the CM peak is angle
independent at around $|\omega_2| \approx 70 $ meV and the IC
peak position $|\omega_1|$ increases from $\approx 25$ to 37 meV
as the angle changes from the nodal to antinodal direction for OP
LSCO.

As alluded in the introduction, this angle dependence of the peak
energy may not be properly addressed in the approaches where a
separable form of the off-diagonal self energy like
$\phi(\vk,\omega)=\phi(\omega)[\cos(k_x a)-\cos(k_y a)]$ or
$\phi(\vk,\omega)=\phi(\omega)\cos(2\theta)$,\cite{Sandvik04prb,Millis92prb,Dahm09naturephys}
or a phenomenological form of the spin susceptibility is assumed.
For example, the separable form of the off-diagonal self-energy
may give misleading results with regard to the angle dependence
of an energy scale because the angle dependence was built in by
hands. No such assumptions were made in this work.

The present results may be checked against the proposed spin
fluctuation theory for the cuprate superconductivity. The most
direct evidence of the spin fluctuation theory will be to detect
the angle and frequency dependence of the diagonal and
off-diagonal self-energy experimentally and compare it with the
results presented here. This will be an extension of the
McMillan-Rowell procedure of phonon
superconductors\cite{McMillan65prl} to $d$-wave pairing. The
experiment of choice for this purpose will be the ARPES because
of its high momentum and frequency resolution capability.

Indeed, the self-energy has been deduced by performing the
momentum distribution curve analysis of the ARPES intensity for
\Bi2212 . For \LSCO \ crystals high quality ARPES data are not
available for comparison though. The requirement of high
resolution ARPES intensity data is being met only recently for
\Bi2212 \ using the Laser
ARPES.\cite{Zhang12prb,Yun11prb,He12arxiv}

If we compare the present calculations on LSCO with the ARPES
analysis from BSCO, bearing the difference in mind, we notice
that the peak intensity of the real part of the self-energy from
BSCO does not change so much as the present calculations. From
the nodal ($\theta=0$) to 30 deg, the peak height changes by less
than a factor of 1.5. The angle dependence of the peak position
is different too. From nodal to anti-nodal direction the peak
position decreases in contrary to the present calculations based
on the AF fluctuations.\cite{Yun11prb,He12arxiv} Proper
comparison, of course, must await high resolution ARPES data from
\LSCO \ materials which may be quantitatively checked against the
present calculations.

%\begin{acknowledgments}

We wish to thank Stephen Hayden for sending us the Ref.\
\cite{Lipscombe08thesis} and Chandra Varma for useful comments on
the manuscript. This work was supported by National Research
Foundation (NRF) of Korea through Grant No.\ NRF 2010-0010772.

%\end{acknowledgments}

%%%%%%%%%%%%%%%%%%%%%       ref          %%%%%%%%%%%%%%%%%%%%%

\bibliographystyle{unsrt}
\bibliography{review}

\end{document}